\documentstyle[11pt,newpasp,twoside,psfig]{article}
\def\edcomment#1{\iffalse\marginpar{\raggedright\sl#1\/}\else\relax\fi}
\reversemarginpar

\begin{document}
\title{Molecular QSO Absorption Line Systems}
 \author{C.~L. Carilli }
\affil{National Radio Astronomy Observatory, Socorro, NM, USA 87801}
\author{K.~M. Menten}
\affil{Max-Planck-Institut f\"ur Radioastronomie, Auf dem H\"ugel 69,
D-53121 Bonn, Germany}

\begin{abstract}

We review observations of molecular absorption line systems at high
redshift toward red quasars and gravitational lenses.

\end{abstract}


The steeply falling power law column density distribution function of 
QSO absorption line systems implies that for every 1000
Ly~$\alpha$ forest lines from absorbers with hydrogen column 
density, $N_{\rm H} < 10^{15}$
cm$^{-2}$ there will be only one system with $N_{\rm H} \ge 10^{22}$
cm$^{-2}$ (Hu et al. 1995). Although rare, these extreme high column
density systems provide the most direct and detailed 
probes of the dense, pre-star-forming interstellar medium (ISM) 
in galaxies
at substantial look-back times. Unfortunately, such high column
densities also lead to substantial optical dust extinction, 
e.g. for a normal dust-to-gas ratio, $A_V$ $> 6$ for $N_{\rm H} > 10^{22}$
cm$^{-2}$;
hence finding such absorbers using optical spectroscopy is
problematic. One method used for finding these systems is
to search for molecular and HI 21cm absorption toward 
flat spectrum radio sources with red (or absent) optical counterparts.
Four high redshift molecular absorption line 
systems have been discovered this way, in two of which 
the absorption occurs in a gravitational lens (0218+357 at $z = 0.685$
and 1830$-$211 at $z = 0.886$; Wiklind \&\ Combes 1995, 1996a).
In the other two systems (1413+135 at $z = 0.247$ and 1504+357 at $z = 0.673$)
the absorption takes place in the host galaxy of an AGN 
(Wiklind \&\ Combes 1994, 1996b).

So far, 15 different molecules have been identified in these redshifted
systems, plus many of their isotopomers, including complex,
multi-atom molecules such as the cyclic species C$_3$H$_2$
(Combes \&\ Wiklind 1999; Menten et al. 1999).
The lines can be relatively broad, as
for 1504+377 with a FWHM = 100 km s$^{-1}$, 
or extremely narrow, as for 1413+135 (FWHM = 1 km
s$^{-1}$). As opposed to emission studies, observations of 
redshifted absorption provide the ability to probe very narrow pencil 
beams through the intervening galaxy and, thus, given a sufficiently strong 
background source, to determine the physical and chemical conditions
in single molecular clouds by observations of rare molecular species.


The best studied of the high-$z$ molecular absorption line sources in terms
of high resolution imaging is that associated with 
the gravitational lens system
1830$-$211 (Wiklind \&\ Combes 1996a). At its redshift of 0.89, many of 
the ground level rotational
transitions of molecules commonly found in galactic molecular clouds
redshift into the
observing bands of the VLA and the VLBA (Carilli et al. 1997, 1998;
Menten et al. 1999; see Fig. 1). Strong absorption is seen toward the 
SW radio component at $z = 0.88582$; e.g, the HCN $J = $ 1--0 line has a 
peak opacity of 2.5 and FWHM = 25 km s$^{-1}$. 
The limit of 0.3 to the opacity 
toward the ``tail'' of the SW component implies an upper limit to the
cloud size of 800 pc for
$H_0$ = 75 km s$^{-1}$ Mpc$^{-1}$, $q_0 = 0.5$.
A VLBA image of the SW component at 24 GHz 
shows a  core-jet structure extending $\approx$ 2 mas to the northwest. 
Spectra  of redshifted  HC$_3$N $J = $ 5--4 absorption at this resolution 
imply a lower limit to the cloud size of 2.5 mas, corresponding to
13 pc, although  there may be sub-structure on scales of 
a few pc (Carilli et al. 1997, 1998). The implied upper limit to the
volume averaged density is 1000 cm$^{-3}$, and the lower limit to the
molecular mass is 3$\times$10$^{4}$ M$_\odot$. 
This is consistent with a lower limit to the cloud size 
of 0.3 pc set by the fact that the line excitation temperatures are
comparable to the microwave background (Frye et
al. 1997). 

\begin{figure}
\vskip -1in
\hspace*{1in}
\psfig{figure=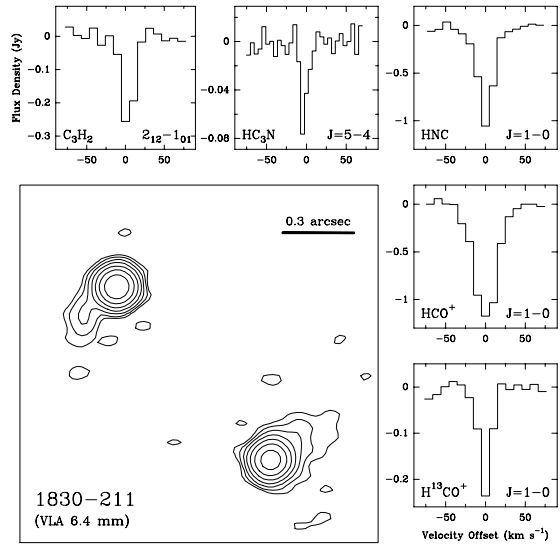,width=10cm}
\vskip -1.5in
\caption{The contour plot shows an image of the
`Einstein ring' radio source PKS 1830$-$211 at
43 GHz with a resolution of 0.1$''$ made with the Very Large Array.
The spectra show molecular absorption by gas in the
lensing galaxy, as observed with the Very Large Array toward the
southwest radio component.
Zero velocity corresponds to a heliocentric redshift
of 0.88582.
}
\end{figure}

Toward the NE component of
1830$-$211 weak molecular absorption is 
detected at a velocity of $-$146 km s$^{-1}$, where zero velocity is
defined by the redshift ($z = 0.88582$) of the absorption seen toward the
SW component. Here the HCN (1--0) optical depth is $\approx 50$ times  
smaller than for the zero velocity component observed toward the SW image 
(Wiklind \&\ Combes 1998; Carilli et al. 1998). 
Chengalur et al. (1999) 
have detected both HI and OH absorption toward
1830$-$211. Interestingly, the HI absorption is stronger at $-$146 km
s$^{-1}$ than at 0 km s$^{-1}$, implying an
[HI/HCN] ratio differing by a factor of 125 between the 
two lines-of-sight, assuming equal spin temperature. 

Menten et al. (1999) conclude that the
molecular abundance ratios for the 1830$-$211 absorber are similar to values
found for Galactic dark clouds and do not show the high abundances of 
HCO$^+$, HCN, and H$_2$CO relative to CO found in many diffuse clouds
(Liszt \&\ Lucas 1995).
Moreover, they find [C$^{12}$/C$^{13}$] = 35, which is smaller by 
a factor of two than the value measured
in the solar system and the local ISM, 
but comparable to that found in some inner Galaxy clouds. 
Since $^{13}$C is only produced in low and intermediate mass stars, while
$^{12}$C is also produced in massive stars, this
ratio is expected to decrease in time and with increasing stellar
processing
(Wilson \&\ Matteucci 1992). 
Studying the deuterium chemistry of the 1830$-$211 absorber,
Shah et al. (1999) set an upper limit of 0.003 on the
[DCN/HCN] ratio.
This is in the upper range of values seen in Galactic dense clouds, 
where the deuterium abundance in the molecular phase is greatly enhanced
due to fractionation. 
Gerin \&\ Roueff (1999) state that this DCN
upper limit is barely compatible with 
a high gas phase deuterium abundance of 10$^{-4}$, perhaps indicating
significant astration (Shah et al. 1999).
A review of oxygen chemistry in 0218+357 is given  in 
Combes \&\ Wiklind (1999). 

Overall, from detailed chemical modeling Gerin \&\ Roueff (1999) 
conclude that the high column density, low density, low O$_2$ and DCN
abundance, and high CCH abundance can be understood if 
the absorbing material is
chemically in the High Ionization Phase (HIP),
in which the fractional ionization is $\sim
10^{-6}$, the gas phase C abundance is high, 
and the chemistry is driven by charge transfer reactions with 
H$^+$. The HIP is the sole stable phase for molecular clouds at low 
densities ($< 10^{4}$ cm$^{-3}$). 


The two known absorption systems associated with lensing galaxies in
gravitational lens systems can be used to study the lens parameters. 
Wiklind \&\ Combes (1998) and Combes \&\ Wiklind (1999) discuss the 
potential of using
single dish observations of the variation of 
molecular absorption line optical depths to measure the geometric 
time delay for 1830$-$211.

The observed difference in
absorption velocities between the SW and NE images in 1830$-$211
allows for tight constraints on the lens mass
distribution and, thus, lensing models.
Wiklind \&\ Combes (1998) derive $V_0 \approx 220 \sqrt{D}$ km s$^{-1}$ for 
the rotation velocity, $V_0$, of the $z \approx 0.89$ lensing galaxy,
with the ``effective distance'' $D$,  in Gpc, in standard
notation.  Using the source redshift of  2.507 
yields $V_0 \approx 366$ km s$^{-1}$, implying a massive
early-type spiral lens.   
The centroid position, inclination angle, and orientation of 
the lens in this model are consistent with the observed values in 
HST near-IR imaging (Lehar et al. 2000).

An interesting use of redshifted molecular absorption lines is to
constrain the evolution of the temperature of the microwave
background radiation. The critical densities for collisional excitation
of the lower rotational lines of high dipole moment molecules 
such as HC$_3$N are typically $\ge
10^5$ cm$^{-3}$. As the absorbing clouds seen toward  
1830$-$211 are likely to have a lower density than this, one expects the
molecular excitation for such molecules to be  
determined by the
ambient radiation field, which, at $z = 0.9$, will be dominated by the
microwave background radiation, at least outside of active star
forming regions. Combes \&\ Wiklind (1999) summarize single dish
measurements of the excitation temperature for a number of molecules
at $z = 0.88582$ toward 1830$-$211, while Menten et al. (1999) present
perhaps the most accurate measurement based on 
VLA observations of the $J = $ 3--2 and 5--4 transitions of HC$_3$N, 
from which they
derive $T_{ex} = 4.5_{-0.6}^{+1.5}$. This is consistent with the
expected microwave background temperature of 5.14 K. 

Drinkwater et al. (1998) and Wiklind \&\ Combes (1998) show how
a comparison of redshifts derived from molecular absorption 
to those derived from  HI 21cm absorption  constrains the cosmic
evolution of the fine structure 
constant, $\alpha$. The current  
limits are not set by the accuracy of the measurements but
by (possible) relative systematic motions of molecular and atomic
absorbing clouds in galaxies. 
Most recently, Carilli et al. (2000) have used VLBI observations
to constrain sight-lines to sub-kpc scales, thereby removing
at least one of the uncertainties in the problem. 
They set a limit of $|{{\dot\alpha}\over{\alpha}}| < 3.5
\times 10^{-15}$~year$^{-1}$ to a look-back time of 4.8 Gyr,
assuming that the velocity error is dictated by small scale ISM
motions. 

{\acknowledgments 
The National Radio Astronomy Observatory is a facility of the National
Science Foundation, operated under  cooperative agreement by
Associated Universities, Inc.}


\begin{references}

\reference Carilli, C.L. et al. 1997, in {\it Structure and Evolution of
the IGM}, eds. Petitjean and Charlot 
(Editions Frontiers: Paris), p. 325
\reference Carilli, C.L. et al. 1998, in {Radio Emission from Galactic 
and Extragalactic Compact Sources,} eds. Zensus et al. (ASP: San
Francisco), p. 317 
\reference Carilli, C.L. et al. 2000, Phys. Rev. Lett., in press
\reference Chengalur, J.N., de Bruyn, A.G., and Narasimha, D. 1999,
A\&A, 343, L79
\reference Combes, F., Wiklind T. 1999, in {\sl Highly Redshifted
Radio Lines}, eds. Carilli et al.
(ASP: San Francisco), p. 210
\reference Drinkwater, M.J. et al. 1998, MNRAS, 295, 457
\reference Frye B., Welch W. J., Broadhurst T.: 1997, ApJ 478, L25
\reference Gerin, M., Roueff, E. 1999, in {\sl
Highly Redshifted Radio Lines}, eds. Carilli, et al.
(ASP: San Francisco), p. 196
\reference Hu, E.M. et al. 1995, AJ, 110, 1526
\reference Lehar, J. et al. 2000 ApJ, 536, 584
\reference Liszt H.S., Lucas R. 1995, A\&A, 299, 847
\reference Menten, K.M., Carilli, C.L., and Reid, M.J. 1999, in {\sl
Highly Redshifted Radio Lines}, eds. Carilli et al. 
(ASP: San Francisco), p. 218
\reference Shah, R. et al. 1999, in {\sl
Highly Redshifted Radio Lines}, eds. Carilli et al.
(ASP: San Francisco), p. 233
\reference Wiklind, T., Combes, F. 1994, A\&A, 286, L9
\reference Wiklind, T., Combes, F. 1995, A\&A, 299, 382
\reference Wiklind, T., Combes, F. 1996a, Nature, 379, 139 
\reference Wiklind, T., Combes, F. 1996b, A\&A, 315, 86 
\reference Wiklind, T., Combes, F. 1998, ApJ, 500, 129
\reference Wilson, T. L., Matteucci, F. 1992, A\&AR, 4, 1
\end{references}
\end{document}